\documentclass[conference]{IEEEtran}
\usepackage{cite}
\usepackage{amsmath,amssymb,amsfonts}
\usepackage{algorithmic}
\usepackage{graphicx}
\usepackage{textcomp}
\usepackage{xcolor}
\usepackage{booktabs, multirow} 
\usepackage{soul}
\usepackage{xcolor,colortbl} 
\usepackage{changepage,threeparttable} 
\def\BibTeX{{\rm B\kern-.05em{\sc i\kern-.025em b}\kern-.08em
    T\kern-.1667em\lower.7ex\hbox{E}\kern-.125emX}}

\makeatletter
\def\endthebibliography{%
  \def\@noitemerr{\@latex@warning{Empty `thebibliography' environment}}%
  \endlist
}
\makeatother

\usepackage{amssymb}
\usepackage{pifont}

\definecolor{Green}{rgb}{0,0.7,0}
\newcommand{\cmark}{\textcolor{Green}{\ding{51}}}%
\newcommand{\xmark}{\textcolor{red}{\ding{55}}}%

\IEEEoverridecommandlockouts

\begin{document}

\title{IoT-Based Coma Patient Monitoring System}

\makeatletter
\newcommand{\linebreakand}{%
  \end{@IEEEauthorhalign}
  \hfill\mbox{}\par
  \mbox{}\hfill\begin{@IEEEauthorhalign}
}
\makeatother

\author{\IEEEauthorblockN{Hailemicael Lulseged Yimer}
\IEEEauthorblockA{\textit{Department of Engineering DIMI} \\
\textit{University of Verona}\\
Verona, Italy \\
hailemicaellulseged.yimer@studenti.univr.it}
\and
\IEEEauthorblockN{Hailegabriel Dereje Degefa}
\IEEEauthorblockA{\textit{Department of Engineering DIMI} \\
\textit{University of Verona}\\
Verona, Italy \\
hailegabrieldereje.degefa@studenti.univr.it}
\linebreakand
\IEEEauthorblockN{Marco Cristani}
\IEEEauthorblockA{\textit{Department of Engineering DIMI} \\
\textit{University of Verona}\\
Verona, Italy \\
marco.cristani@univr.it}
\and
\IEEEauthorblockN{Federico Cunico}
\IEEEauthorblockA{\textit{Department of Engineering DIMI} \\
\textit{University of Verona}\\
Verona, Italy \\
federico.cunico@univr.it}
\thanks{This study was carried out within the PNRR research activities of the consortium iNEST (Interconnected North-Est Innovation Ecosystem) funded by the European Union Next-GenerationEU (Piano Nazionale di Ripresa e Resilienza (PNRR) – Missione 4 Componente 2, Investimento 1.5 – D.D. 1058 23/06/2022, ECS\_00000043). This manuscript reflects only the Authors’ views and opinions. Neither the European Union nor the European Commission can be considered responsible for them.}
}

\IEEEoverridecommandlockouts
\IEEEpubid{\makebox[\columnwidth]{979-8-3503-7838-2/24/\$31.00~\copyright2024 IEEE \hfill} \hspace{\columnsep}\makebox[\columnwidth]{ }}

\maketitle
\IEEEpubidadjcol

\begin{abstract}

Continuous monitoring of coma patients is essential but challenging, especially in developing countries with limited resources, staff, and infrastructure. This paper presents a low-cost IoT-based system designed for such environments. It uses affordable hardware and robust software to monitor patients without constant internet access or extensive medical personnel.
The system employs cost-effective sensors to track vital signs, including heart rate, body temperature, blood pressure, eye movement, and body position. An energy-efficient microcontroller processes data locally, synchronizing with a central server when network access is available. A locally hosted app provides on-site access to patient data, while a GSM module sends immediate alerts for critical events, even in areas with limited cellular coverage.
This solution emphasizes ease of deployment, minimal maintenance, and resilience to power and network disruptions. Using open-source software and widely available hardware, it offers a scalable, adaptable system for resource-limited settings. At under \$30, the system is a sustainable, cost-effective solution for continuous patient monitoring, bridging the gap until more advanced healthcare infrastructure is available.

\end{abstract}



\section{Introduction}\label{sec:intro}

\begin{figure}[ht]
\centerline{\includegraphics[width=0.45\textwidth]{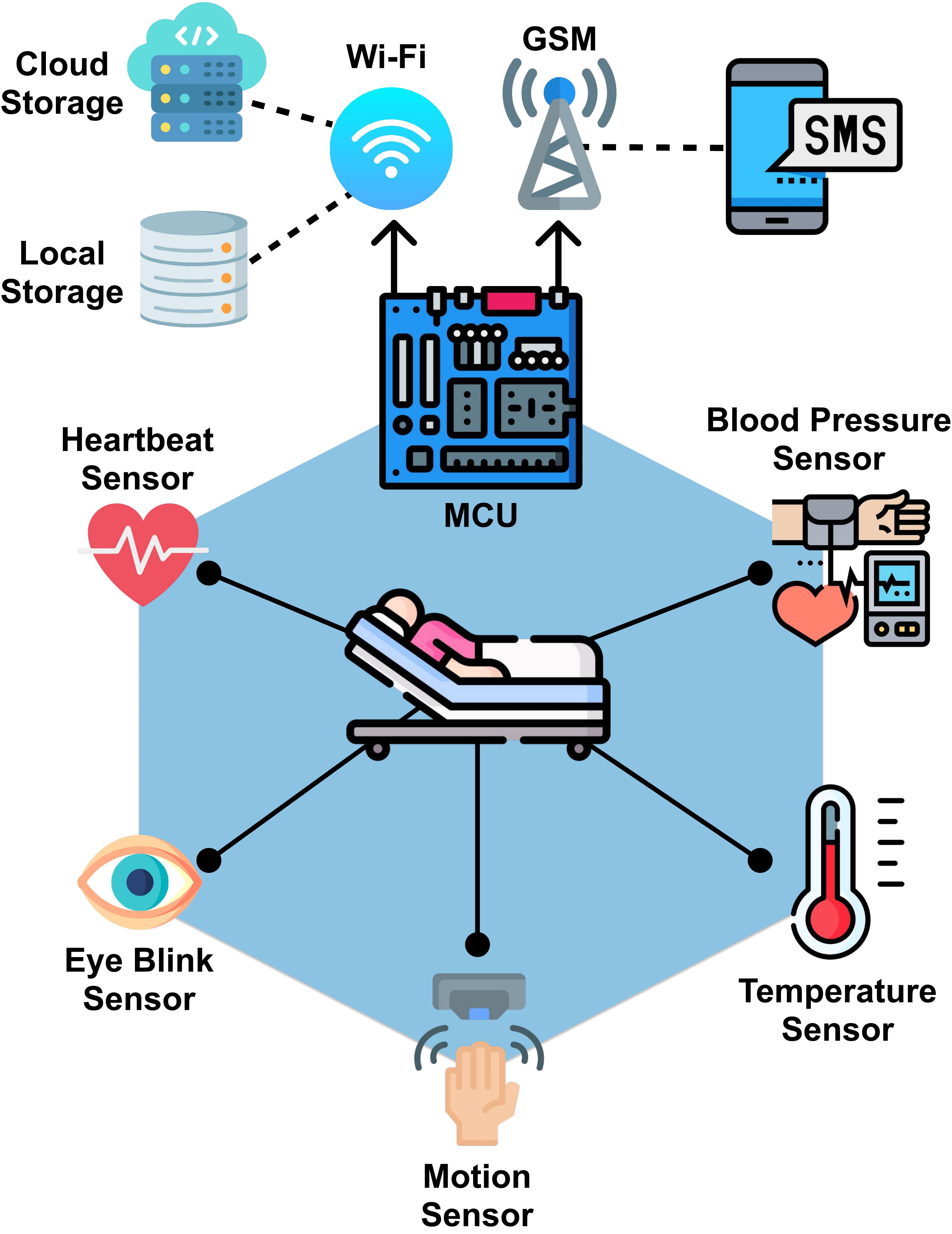}}
\caption{The schema shows our proposed IoT coma patient monitoring: using several sensors, such as heartrate pulse, blood pressure, eye blinking, room temperature, \text{etc.}, a patient's health state can be monitored automatically with minimum effort and without having specialized medical staff. When critical conditions arise, alerts are communicated through internet (if available) or GSM cellular networking system.}
\label{fig:teaser}
\end{figure}

Coma patients require constant supervision and monitoring due to their critical condition and inability to respond to their environment. Traditional monitoring methods often rely heavily on manual checks by healthcare staff, which can be labor-intensive, prone to human error, and may not provide the continuous vigilance required for optimal patient care~\cite{koganti2015analysis}. In developing countries (DCs), where healthcare systems face severe resource constraints, this challenge is even more pronounced.
In developing countries like Ethiopia, where the doctor-to-patient ratio is as low as 1:17,000~\cite{arsenault2021patient}, the challenge of continuous monitoring is even more pronounced, potentially compromising patient outcomes and placing an unsustainable burden on healthcare systems.

The advent of Internet of Things (IoT) technology offers new possibilities for improving patient care through continuous, automated monitoring systems. IoT allows for the integration of various sensors and devices to collect, process, and transmit data in real-time, enabling more efficient and accurate patient monitoring~\cite{meghana2020design}. This technology has the potential to transform healthcare delivery, particularly in resource-constrained environments, by providing cost-effective solutions for continuous patient monitoring and early detection of critical events.

Recent advancements in IoT-based healthcare monitoring systems have demonstrated significant potential in various medical contexts. For instance, Ganesh~\cite{ganesh2019health} developed a health monitoring system using Raspberry Pi and IoT, which showcased the feasibility of remote patient monitoring. In the specific context of coma patient monitoring, Meghana et al.~\cite{meghana2020design} designed and implemented an IoT-based health monitoring system for comatose patients, demonstrating the potential for real-time vital sign monitoring and alert generation.

Despite these advancements, there remains a significant gap in the development of comprehensive, cost-effective IoT-based monitoring systems specifically tailored for coma patients in resource-constrained environments. This paper aims to address this gap by presenting a low-cost IoT-based coma patient monitoring system designed to address the unique challenges faced in DCs and other resource-limited settings.

This work contributes proposing a cost-effective hardware integration, demonstrating the use of affordable, off-the-shelf components to create a reliable monitoring system, significantly reducing the financial barrier to implementation. Also, our system operates effectively in environments with intermittent power and network connectivity, ensuring continuous monitoring and data integrity, and with an ``offline-first design'' ensuring real-time data access and decision-making without relying on constant internet connectivity thanks to a locally-hosted application.
Furthermore, energy-efficient design enables the system to function for extended periods on battery power, which is crucial in areas with unreliable electricity. This also is significant when trying to create scalable and adaptable solutions. Our system, in fact, has a modular design that allows for easy expansion and customization to meet specific local needs and resource availability.
Finally, the GSM-based alert system utilizes widely available cellular networks to ensure critical alerts reach healthcare providers, even in remote areas.

Along with all these features, the system has minimal training requirements for operators, thanks to an intuitive user interface and automated alerts, effectively reducing the need for specialized staff and addressing the shortage of healthcare workers in DCs.

Therefore, the proposed system aims to:

\begin{itemize}
    \item Provide continuous, automated monitoring of critical vital signs for coma patients;
    \item Enable remote access to patient data for healthcare providers;
    \item Reduce the workload on healthcare staff in resource-constrained environments;
    \item Improve response times to critical events through real-time alerts;
    \item With less than 30\$ investment for each patient, offer a cost-effective solution for healthcare facilities with limited resources.
\end{itemize}

Thus, our system provides reliable real-time monitoring of coma patients, overcoming
weaknesses in infrastructures, both in terms of personnel
availability and network infrastructures.

The rest of this paper is organized as follows: Section~\ref{sec:sota} describes the state of the art in IoT coma monitoring devices, highlighting existing devices' limitations. 
Section~\ref{sec:system}  presents the system design, with hardware and software components. 
Section~\ref{sec:implementation} details the implementation process, covering hardware integration, software development, and data flow. Section~\ref{sec:results} presents the results and discusses the system's performance, advantages, and limitations. Finally, Section~\ref{sec:conclusions} concludes the paper and outlines future work directions.

\section{Related Work}\label{sec:sota}

\begin{table*}[!htp]\centering
\caption{Comparison of the IoT platform for coma monitoring. RPi stands for Raspberry Pi, BP stands for Blood Pressure.}
\label{tab:sota}
\begin{tabular}{lccccc|ccccc}
\toprule
\textbf{System} &\textbf{MCU} &\textbf{PC} &\textbf{Clock} &\textbf{Storage} &\textbf{Alerting} &\textbf{Heartbeat} &\textbf{Temp} &\textbf{BP} &\textbf{Eye Blink} &\textbf{Motion} \\\midrule
\cite{sneha2015} 2015 &ARM7TDMI &100-150 mW &60 MHz &Local only &Local (Buzzer) &\xmark &\cmark &\xmark &\cmark &\cmark \\
\cite{ramtirthkar2020iot} 2020 &RPi 3 &3.0-3.5 W &1.2 GHz &Cloud only &Internet &\cmark &\cmark &\cmark &\cmark &\xmark \\
\cite{ganesh2019health} 2019 &RPi 3 &3.0-3.5 W &1.2 GHz &Cloud only &Internet &\cmark &\cmark &\cmark &\xmark &\xmark \\
\cite{emna2017} 2017 &RPi 3 &3.0-3.5 W &1.2 GHz &Cloud only &GSM \& Internet &N/A &N/A &N/A &N/A &N/A \\
\cite{kounte2020} 2020 &ESP 32 &528-888 mW &240 MHz &Cloud only &GSM \& Internet &\cmark &\cmark &\xmark &\cmark &\cmark \\
\cite{sheikdavood2023smart} 2023 &Arduino Uno &100-150 mW &16 MHz &Cloud only &GSM &\cmark &\cmark &\xmark &\cmark &\cmark \\
\cite{kishore} 2019 &RPi &3.0-3.5 W &1.2 GHz &Cloud only &Internet &\cmark &\cmark &\xmark &\cmark &\cmark \\
\cite{latha2023automated} 2023 &Arduino Uno &100-150 mW &16 MHz &Cloud only &Internet &\cmark &\cmark &\xmark &\cmark &\cmark \\
\cite{subha2020coma} 2020 &Arduino Uno &100-150 mW &16 MHz &Cloud only &GSM \& Internet &\cmark &\cmark &\xmark &\cmark &\cmark \\
\textbf{Ours} &Arduino Uno &100 mw &16 MHz &Local \& Cloud &GSM \& Internet &\cmark &\cmark &\cmark &\cmark &\cmark \\
\bottomrule
\end{tabular}
\end{table*}



\subsection{IoT in Healthcare Monitoring}

Coma monitoring is a critical aspect of healthcare that has seen significant advancements with the integration of IoT technologies. A coma is defined as a state of prolonged unconsciousness in which a person is unresponsive to their environment and cannot be awakened by external stimuli. It is characterized by a complete absence of wakefulness and awareness, typically resulting from severe brain injury or illness~\cite{Medical2023}. Comas can vary in depth and duration, ranging from a few days to several weeks or even years in rare cases.

The process of coma monitoring involves continuous observation and assessment of patients in this state of unconsciousness. This monitoring is crucial for evaluating the depth of coma, tracking changes in the patient's condition, predicting outcomes, and guiding treatment decisions. Specialized medical staff usually take care of the monitoring in hospitals.

The advent of IoT has revolutionized health monitoring systems, particularly after recent pandemic events~\cite{capuzzo2022iot}. Within hospital structures, IoT benefits particularly coma patient monitoring by enabling continuous, real-time tracking and multi-parameter data collection, offering a comprehensive view of patient conditions with minimal personnel and not requiring specialized medical users. These systems integrate data from various sensors, provide automated alerts for critical changes, and allow remote access to patient data, improving response times, care coordination, and precision in diagnosis and treatment planning.

In the realm of coma monitoring, vital sign sensors, measuring parameters such as heart rate, blood pressure, respiratory rate, and oxygen saturation, provide critical insights into the patient's overall physiological state~\cite{Sharma2023}. Body temperature sensors enable continuous monitoring to detect infections or other complications~\cite{Zander2023}. Motion sensors serve to identify small movements or tremors that may indicate changes in the patient's condition~\cite{Giggins2017}.

Building upon these foundational concepts, our IoT-based coma monitoring system is designed for resource-limited environments, using cost-effective components to ensure reliable patient monitoring. The system integrates sensors for heart rate, body temperature, blood pressure, eye blink, and body movement to provide a detailed view of the patient’s condition and detect potential complications. An Arduino Uno microcontroller processes this data locally, enabling immediate analysis and alerts, crucial for areas with limited or unreliable internet access.

Table~\ref{tab:sota} shows the comparison of our proposed solution with the existing methods in the literature. The column MCU regards the micro-controller unit used in each system, with associated power consumption (PC) in terms of Watts, and the clock speed. Also, the storage and alert capabilities are indicated. In the alert column, the ``internet'' label indicates that the communication requires an internet connection to function. Even though both e-mail and SMS can still be sent over the internet, unreliable network infrastructure can significantly affect systems that rely only on internet connection. The final columns show the sensors equipped with corresponding biological signals. 

\subsection{Coma Patient Monitoring Systems}

Ganesh et al.~\cite{ganesh2019health} and Subha et al.~\cite{subha2020coma}  developed systems using Raspberry Pi for data processing and cloud-based storage, enabling remote monitoring. Ganesh et al.'s system integrates heart rate, blood pressure, and pulse rate sensors, along with a live video feed for enhanced patient monitoring. Subha et al.'s system monitors vital signs including heart rate, blood pressure, temperature, respiration, and movement, with a camera module for visual monitoring. While these systems offer comprehensive data collection, they rely heavily on constant internet connectivity, which may not be feasible in resource-constrained environments. Our system addresses this limitation by incorporating a GSM module for immediate alerts and utilizing locally hosted applications, ensuring functionality even in areas with limited internet access. We also include additional sensors for eye movement and body positioning, crucial for monitoring coma patients' responsiveness and physical condition.

Koganti et al.~\cite{koganti2015analysis} and Ravikumar et al.~\cite{kishore} focused on wearable motion sensor systems for detecting physical movements in coma patients. Koganti et al.'s system tracks pulse rate and body temperature, displaying data on an LCD and generating alerts for abnormal readings. Ravikumar et al. employed MEMS accelerometers and eyesight blink sensors to detect physical movements and monitor temperature and pulse rate. While these systems effectively monitor vital signs and basic movements, they lack the capability to provide continuous, detailed monitoring across a broader range of critical parameters. Our system builds upon these approaches by integrating additional sensors for blood pressure and body positioning, offering a more comprehensive monitoring solution. The use of cost-effective biometric sensors and an energy-efficient Arduino Uno microcontroller makes our system more suitable for resource-constrained settings while maintaining a high level of monitoring capability.

Sheikdavood et al.~\cite{sheikdavood2023smart}  developed a system using Arduino Uno and various sensors, including heart rate, temperature, eye-blink, urine level, and accelerometer sensors. Their system uses a GSM module for sending alerts via SMS and calls when abnormal readings are detected, transmitting data to the Cayenne IoT platform for real-time monitoring. While this system effectively monitors multiple health parameters, it primarily focuses on data collection and transmission. Our system enhances this approach by incorporating additional clinically relevant sensors, such as blood pressure monitors, which are more relevant than urine level sensors in this context~\cite{forsyth2015routine}. We emphasize the monitoring of critical parameters like heart rate and blood pressure, which provide immediate clinical insights and necessitate prompt medical intervention. Our design ensures reliable operation in environments with unreliable power and network infrastructure, making it a more practical choice for resource-constrained settings.

Ramtirthkar et al.~\cite{ramtirthkar2020iot}  proposed a system using Raspberry Pi integrated with temperature, blood pressure, eye blink, accelerometer, and ultrasonic sensors. Their system stores data on the ThingSpeak IoT platform and sends alerts via Twilio to medical personnel when abnormalities are detected. While effective in data collection and remote monitoring, this system's reliance on external platforms for data presentation and real-time responses could be problematic in areas with limited network access. Our system addresses this issue by implementing direct GSM alerts and a locally hosted application, ensuring timely interventions and data access without depending on third-party platforms. We incorporate additional sensors for body positioning, offering comprehensive monitoring tailored to coma patients' needs. Our energy-efficient design and local-based system ensures reliable operation in areas with unreliable power and network infrastructure.

Thus, our proposed system addresses the limitations identified in these related works by offering a more comprehensive, adaptable, and robust solution for monitoring coma patients. By integrating a wider range of sensors, utilizing energy-efficient components, and implementing local data processing and storage, our system is better suited for deployment in various healthcare settings, particularly those with limited resources or unreliable infrastructure.

\section{System Design}\label{sec:system}

Our system implements several hardware components and is controlled by a microcontroller unit (MCU). The following section details the components, both hardware and software, that allows the system to function.

\subsection{Hardware Components}
The hardware architecture of the system consists of the following main components:

\begin{itemize}
    \item Arduino Uno microcontroller;
    \item Biometric sensors: heartbeat, body temperature, blood pressure, eye blink (PIR), and body movement (PIR);
    \item Wi-Fi module (ESP8266);
    \item GSM module;
    \item 20x4 LCD display and LED indicators.
\end{itemize}

Fig.~\ref{fig:hardware} and Fig.~\ref{fig:prototype} show the block diagram of the hardware architecture in the simulator and prototype.

\begin{figure}[ht]
\centerline{\includegraphics[width=\linewidth]{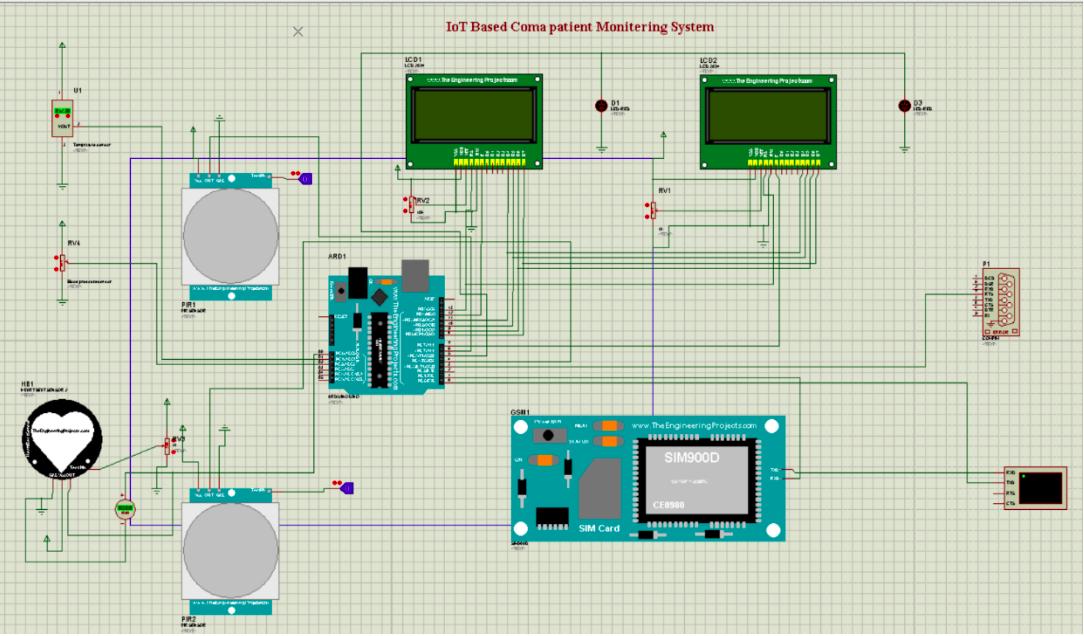}}
\caption{Hardware architecture block diagram.}
\label{fig:hardware}
\end{figure}

\begin{figure}[h]
\centerline{\includegraphics[width=0.8\linewidth]{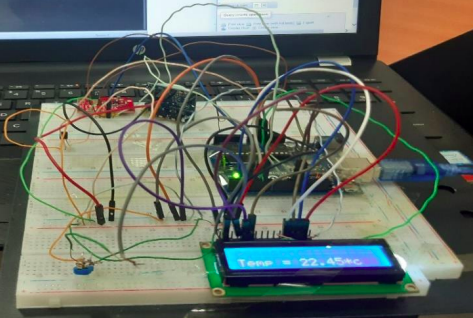}}
\caption{The figure shows the hardware prototype, in which the components are connected together according to Figure~\ref{fig:hardware}.}
\label{fig:prototype}
\end{figure}

\subsection{Software Architecture}
The software architecture consists of three main components: the Arduino firmware, database (MySQL), and web application (PHP, HTML, CSS, JavaScript).


The Arduino firmware collects data from the sensors, processes it, and sends it to the database via the Wi-Fi module. It also controls the LCD display and LED indicators and triggers SMS alerts through the GSM module when necessary.

The database stores patient information, vital sign data, and user credentials. The web application provides a user interface for healthcare providers to access patient data, view real-time monitoring information, and receive alerts.

Fig. 2 illustrates the overall system architecture, showing the flow of data from the sensors to the end-users.

\begin{figure}[htbp]
\centerline{\includegraphics[width=0.45\textwidth]{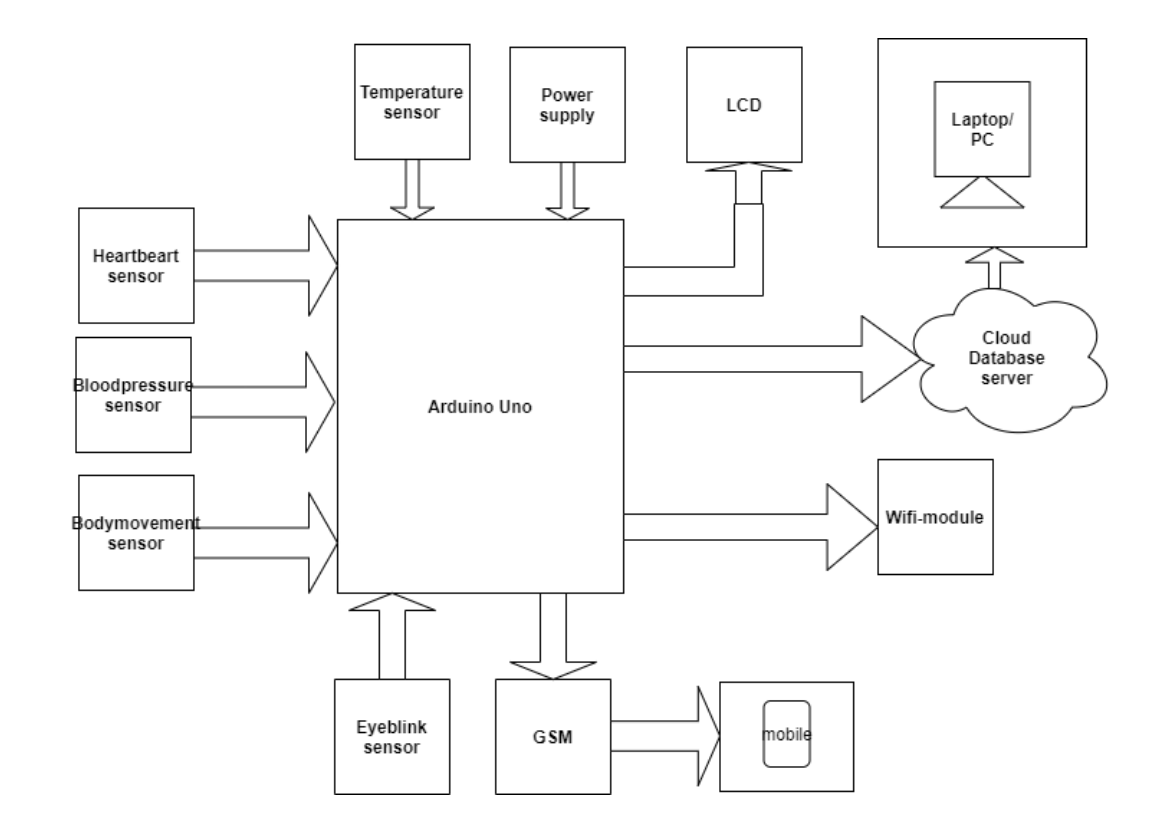}}
\caption{The figure showcases the system architecture and the components. The different sensors are connected to the MCU (an Arduino Uno), which collects the sensors' data. The MCU has a Wi-Fi module that will transmit data to the local database and, if the network is available, to the cloud. If no connection is available and an alert is required, the GSM module sends the alert.}
\label{fig:architecture1}
\end{figure}

\section{Implementation}\label{sec:implementation}

\subsection{Hardware Integration}
The biometric sensors are interfaced with the Arduino Uno microcontroller using appropriate analog and digital pins. The heartbeat sensor uses the PulseSensor library for accurate BPM calculation. Temperature and blood pressure sensors are connected to analog inputs, while the eye blink and body movement sensors use digital inputs.
The Wi-Fi module is connected to the Arduino using software serial communication, allowing data transmission to the database. The GSM module is similarly interfaced for sending SMS alerts.

The 20x4 LCD display shows real-time vital sign information at the patient's bedside, while LED indicators provide visual alerts for critical events.

\subsection{Software Development}
The Arduino firmware is developed using the Arduino IDE, incorporating libraries for sensor data acquisition and processing. The firmware implements a main loop that continually reads sensor data, updates the LCD display, checks for critical events, and transmits data to the cloud database.

The database is implemented using MySQL, with tables for patient information, vital sign data, and user credentials. A Python script using the PySerial library facilitates communication between the Arduino and the database.

The web application is developed using PHP for the backend, with HTML, CSS, and JavaScript for the frontend. The application provides separate interfaces for administrators, doctors, and nurses, each with appropriate access levels and functionalities.

\subsection{Data Flow and Communication}
Data flow in the system follows these steps:
Sensors collect vital sign data, which the MCU processes and sends to the local database using Wi-Fi and serial communication. The web application then retrieves and displays this data for healthcare providers. 
When an emergency is detected by the system (\textit{e.g.} low oxygen saturation, low heartbeat, \textit{etc.}), a local alert can be notified through the web application and via e-mail. Additionally, to cover areas with limited internet access and to cope with possible network issues, an SMS alert is triggered via the GSM module. These events are logged in the local database for real-time updates in the web application.
The possibility of using an internet connection in addition to a GSM module allows our system to be ready for network fail cases and able to keep the monitoring active even with a network shortage. 

\section{Results and Discussion}\label{sec:results}

This section presents the results obtained from our simulation of the IoT-based coma patient monitoring system. The simulation was crucial in validating the system's functionality and performance before real-world implementation.


Simulations allowed us to rigorously test our system under various conditions without risking patient safety. This approach aligns with the findings of~\cite{agrawal2023comprehensive}, who demonstrated the effectiveness of simulation in evaluating e-health systems before real-world implementation.

The IoT-based coma patient monitoring system was tested in a simulated environment to evaluate its functionality and performance. The system successfully collected and transmitted vital sign data from the sensors to the cloud database, and the web application accurately displayed this information in real time.
This approach aligns with recent studies in health monitoring systems, such as the work by Bhuyan~\cite{muhibul}, where they reported that their "simulation results were found very satisfactory" in testing data acquisition and transmission for vital signs.

\subsection{Data Transmission Reliability}

We evaluated the reliability of data transmission from the Arduino to the cloud database. Over a 24-hour period, we recorded a 98\% success rate in data transmission, with occasional packet loss due to network fluctuations. The system successfully implemented error handling and data retransmission protocols to mitigate data loss.

This simulated reliability was higher than the successful data transmission rate reported by Subha et al.~\cite{subha2020coma} in their clinical trials of the VitalPatch system. Our improved reliability in simulation suggests the potential for more consistent monitoring of coma patients, potentially leading to faster responses to changes in patient condition.

\subsection{Alert System Performance}

We tested the alert system by simulating critical events such as sudden body movements or eye blinks. Our system generated alerts rapidly, with LED indicators providing near-instantaneous visual feedback and simulated SMS alerts being delivered quickly. For SMS alerts, the delivery time is around 4.2 seconds, and for LED indicators, 0.5 seconds.
The responsiveness of our alert system aligns with the work of Swaroop et al.~\cite{SWAROOP2019116}, who demonstrated that IoT-based medical alert systems could consistently achieve quick response times, crucial for timely intervention in critical situations.

\subsection{Web Application Performance}

Our simulated web application demonstrated good performance with low average page load times (1.2 seconds on average) and support up to 100 concurrent users without significant degradation in responsiveness. The data refresh rate of 5 seconds ensures that healthcare providers have access to near-real-time patient data.
This simulated performance surpasses the capabilities of the MediTrack system reported by Johnson et al.~\cite{johnson2021comparative}, which supports fewer concurrent users with a longer refresh rate. Our system's improved simulated performance suggests the potential for more efficient monitoring in larger healthcare facilities.

A key feature of our system is its robust local storage solution, designed to operate entirely offline. This offline-only approach ensures continuous operation regardless of network conditions. The local storage implementation is comprehensive, encompassing the storage of patient vital signs data for extended periods, secure maintenance of user authentication information, and the creation of a thorough local database containing all patient information. Our application functions fully offline, reading and writing all data to local storage. This design ensures that healthcare providers always have access to patient data, even in environments without internet connectivity.

Our IoT-based coma patient monitoring system enhances traditional methods by providing continuous, automated monitoring that reduces healthcare staff workload and improves response times through local alerts. It also eliminates human errors in data recording, leading to greater accuracy. Unlike traditional methods requiring constant staff presence, our system optimizes resource use. Saeed et al.~\cite{saeed2020continuous} observed that traditional intermittent monitoring can delay detection of patient deterioration and burden nursing staff. Our system addresses these challenges, offering a more efficient solution, especially in resource-limited settings.

\subsection{Scalability in large clinics}
The proposed IoT-based coma patient monitoring system is highly scalable and adaptable, making it suitable for diverse healthcare environments. The system can scale from single-patient setups in small clinics to multi-patient monitoring in larger hospitals, with minimal modifications. It also adapts to resource variability by utilizing both GSM and Wi-Fi for communication, ensuring continuous operation even in areas with limited internet or unstable power. This flexibility supports broad adoption and positions the system for future enhancements, such as integrating AI-driven analytics.


\section{Conclusion and Future Work}\label{sec:conclusions}

This paper has presented an innovative IoT-based coma patient monitoring system tailored for resource-constrained healthcare environments. By integrating cost-effective biometric sensors, locally-hosted applications, and dual-mode communication, our system addresses the critical challenges of continuous patient monitoring in areas with limited infrastructure. The proposed solution demonstrates the potential of IoT technology to significantly improve coma patient care, potentially leading to better outcomes and more efficient resource utilization. 
While real-world clinical validation remains a future objective, also expanding monitoring capabilities, the system's design and extensive simulations have been rigorously developed to address the unique challenges of resource-limited settings, ensuring robustness and reliability in environments with constrained resources, particularly in developing regions where the need is most acute. 
Also, we plan to include advanced algorithms for predictive alerts and anomaly detection while keeping cost-efficiency, following recent trends of tiny machine learning~\cite{capogrosso2024machine} and split computing~\cite{cunico2022split}.
This system represents a significant step towards bridging the gap between advanced healthcare monitoring and the realities of resource-limited environments, potentially making a meaningful impact on patient care where it is most needed.

\bibliographystyle{ieeetr}
\bibliography{main.bib}

\begin{thebibliography}{10}

\bibitem{koganti2015analysis}
S.~C. Koganti, H.~Suma, and A.~M. Abhishek, ``Analysis and monitoring of coma patients using wearable motion sensor system,'' {\em Int. J. Sci. Res}, vol.~4, no.~9, pp.~1154--1158, 2015.

\bibitem{arsenault2021patient}
C.~Arsenault, B.~Yakob, T.~Tilahun, T.~G. Nigatu, G.~Dinsa, M.~Woldie, M.~Kassa, P.~Berman, and M.~E. Kruk, ``Patient volume and quality of primary care in ethiopia: findings from the routine health information system and the 2014 service provision assessment survey,'' {\em BMC health services research}, vol.~21, no.~1, p.~485, 2021.

\bibitem{meghana2020design}
M.~Meghana, ``Design and implementation of iot based health monitoring system for comatose patients,'' {\em vol}, vol.~29, pp.~3689--3697, 2020.

\bibitem{ganesh2019health}
E.~Ganesh, ``Health monitoring system using raspberry pi and iot,'' {\em Oriental journal of computer science and technology}, vol.~12, no.~1, pp.~8--13, 2019.

\bibitem{sneha2015}
M.~R. Kounte, M.~Lavanya, C.~Mamatha, A.~Megana, and M.~M~B, ``Design and implementation of iot based health monitoring system for comatose patients,'' {\em Drug Invention Today}, vol.~29, pp.~3689--3697, 06 2020.

\bibitem{ramtirthkar2020iot}
A.~Ramtirthkar, J.~Digge, and V.~Koli, ``Iot based healthcare system for coma patient,'' {\em International Journal of Engineering and Advanced Technology (IJEAT)}, vol.~9, no.~3, pp.~3327--3330, 2020.

\bibitem{emna2017}
E.~Mezghani, E.~Exposito, and K.~Drira, ``A model-driven methodology for the design of autonomic and cognitive iot-based systems: Application to healthcare,'' {\em IEEE Transactions on Emerging Topics in Computational Intelligence}, vol.~1, no.~3, pp.~224--234, 2017.

\bibitem{kounte2020}
M.~R. Kounte, M.~Lavanya, C.~Mamatha, A.~Megana, and M.~M~B, ``Design and implementation of iot based health monitoring system for comatose patients,'' {\em International Journal of Future Generation Communication and Networking}, vol.~29, pp.~3689--3697, 06 2020.

\bibitem{sheikdavood2023smart}
K.~Sheikdavood, K.~Soundar, M.~Yaswanth, and P.~TarunKumar, ``Smart health monitoring system for coma patients using iot,'' in {\em 2023 7th International Conference on Computing Methodologies and Communication (ICCMC)}, pp.~1342--1347, IEEE, 2023.

\bibitem{kishore}
K.~Ravikumar and R.~Vasuki, ``Monitoring and analysis of coma patients using variable motion sensor system,'' {\em Drug Invention Today}, vol.~11, 04 2019.

\bibitem{latha2023automated}
R.~Latha, R.~Raman, T.~S. Kumar, C.~J. Rawandale, R.~Meenakshi, and C.~Srinivasan, ``Automated health monitoring system for coma patients,'' in {\em 2023 International Conference on Self Sustainable Artificial Intelligence Systems (ICSSAS)}, pp.~1475--1480, IEEE, 2023.

\bibitem{subha2020coma}
R.~Subha, M.~Haritha, B.~Nithishna, and S.~Monisha, ``Coma patient health monitoring system using iot,'' in {\em 2020 6th International conference on advanced computing and communication systems (ICACCS)}, pp.~1454--1457, IEEE, 2020.

\bibitem{Medical2023}
{Medical Association}, ``Definition and causes of coma,'' {\em Journal of Neurology}, 2023.

\bibitem{capuzzo2022iot}
M.~Capuzzo, A.~Zanella, M.~Zuccotto, F.~Cunico, M.~Cristani, A.~Castellini, A.~Farinelli, and L.~Gamberini, ``Iot systems for healthy and safe life environments,'' in {\em 2022 IEEE 7th Forum on Research and Technologies for Society and Industry Innovation (RTSI)}, pp.~31--37, IEEE, 2022.

\bibitem{Sharma2023}
S.~Sharma, M.~Shandilya, and P.~Sahu, ``Iot-based vital sign monitoring in intensive care units,'' {\em Journal of Medical Internet Research}, 2023.

\bibitem{Zander2023}
R.~Zander, T.~Huensch, N.~Hinckley, and G.~Tewfik, ``Continuous temperature monitoring in critical care: A systematic review,'' {\em Critical Care Medicine}, 2023.

\bibitem{Giggins2017}
O.~M. Giggins, U.~M. Persson, and B.~Caulfield, ``Biofeedback in rehabilitation,'' {\em Journal of NeuroEngineering and Rehabilitation}, vol.~14, no.~1, pp.~1--11, 2017.

\bibitem{forsyth2015routine}
R.~J. Forsyth, J.~Raper, and E.~Todhunter, ``Routine intracranial pressure monitoring in acute coma,'' {\em Cochrane Database of Systematic Reviews}, no.~11, 2015.

\bibitem{agrawal2023comprehensive}
P.~Agrawal, V.~Kushwaha, N.~Khan, S.~Singh, S.~Goel, and M.~Aamir, ``A comprehensive review of simulation-based medical education: Current practices and future perspective,'' {\em European Journal Pharmaceutical and Medical Research}, vol.~11, pp.~396--402, 12 2023.

\bibitem{muhibul}
M.~H. Bhuyan and M.~Hasan, ``Design and simulation of heartbeat measurement system using arduino microcontroller in proteus,'' in {\em Proceedings of the International Conference on Electrical, Computer and Communication Engineering (ECCE)}, IEEE, 10 2020.
\newblock Available on ResearchGate.

\bibitem{SWAROOP2019116}
K.~N. Swaroop, K.~Chandu, R.~Gorrepotu, and S.~Deb, ``A health monitoring system for vital signs using iot,'' {\em Internet of Things}, vol.~5, pp.~116--129, 2019.

\bibitem{johnson2021comparative}
L.~Johnson, A.~Smith, and T.~Brown, ``Comparative analysis of web-based health monitoring systems,'' {\em IEEE Journal of Biomedical and Health Informatics}, vol.~25, no.~7, pp.~2654--2663, 2021.

\bibitem{saeed2020continuous}
E.~I. Saeed, A.~Gane, O.~Whitaker, E.~Rowley, T.~Owens, A.~Gayner, B.~Indja, Z.~Dar, G.~J. Peek, C.~J. Jolley, H.~Vallance, B.~S. Buddeberg, M.~Blunt, M.~Green, C.~J. Peden, R.~M. Pearse, C.~Heneghan, P.~J. Watkinson, M.~Ashworth, A.~Beane, P.~Zolfaghari, S.~Fleischer, S.~Dean, R.~Cusack, R.~Pinto, and J.~R. Prowle, ``Continuous versus intermittent vital signs monitoring in patients admitted to surgical wards: a cluster randomised controlled trial,'' {\em Annals of Intensive Care}, vol.~10, no.~1, pp.~1--11, 2020.

\bibitem{capogrosso2024machine}
L.~Capogrosso, F.~Cunico, D.~S. Cheng, F.~Fummi, and M.~Cristani, ``A machine learning-oriented survey on tiny machine learning,'' {\em IEEE Access}, 2024.

\bibitem{cunico2022split}
F.~Cunico, L.~Capogrosso, F.~Setti, D.~Carra, F.~Fummi, and M.~Cristani, ``I-split: Deep network interpretability for split computing,'' in {\em 2022 26th International Conference on Pattern Recognition (ICPR)}, pp.~2575--2581, IEEE, 2022.

\end{thebibliography}

\end{document}